\documentclass{aa}  
\def\lesssim{\mathrel{\hbox{\rlap{\hbox{\lower4pt\hbox{$\sim$}}}\hbox{$<$}}}}

\usepackage{graphicx}
\usepackage{lscape}
\usepackage{txfonts}
\begin{document}
   \title{X-rays from Saturn: A study with {\it XMM-Newton} and {\it Chandra}
\\
over the years 2002-05}


   \author{G. Branduardi-Raymont
          \inst{1}
          \and
          A. Bhardwaj
          \inst{2}
          \and
          R. F. Elsner
          \inst{3}
          \and
          P. Rodriguez
          \inst{4}           }
   
   \offprints{G. Branduardi-Raymont
              \email{gbr@mssl.ucl.ac.uk}}

   \institute{Mullard Space Science Laboratory, University College London,
              Holmbury St Mary, Dorking, Surrey RH5 6NT, UK
    \and
              Space Physics Laboratory, Vikram Sarabhai
              Space Centre, Trivandrum 695022, India
    \and
              NASA Marshall Space Flight Center, NSSTC/XD12,
              320 Sparkman Drive, Huntsville, AL 35805, USA
    \and
              XMM-Newton SOC, Apartado 50727, Villafranca, 28080 Madrid,
              Spain
             }

   \date{Received ...; accepted ...}

 
  \abstract
   {}
   {We approach the study of Saturn and its environment in a novel way using 
X-ray data, by making a systematic and uniform spectral analysis of all the 
X-ray observations of the planet to date. }
 {We present the results of the two most recent (2005) XMM-Newton observations 
of Saturn together with the re-analysis of an earlier (2002) observation from 
the XMM-Newton archive and of three Chandra observations in 2003 and 2004. 
While 
the XMM-Newton telescope resolution does not enable us to resolve spatially
the contributions of the planet's disk and rings to the X-ray flux, we can 
estimate their strengths and their evolution over the years from spectral 
analysis, and compare them with those observed with Chandra.}
{The spectrum of the X-ray emission is well fitted by an 
optically thin coronal model with an average temperature of 0.5 keV. The 
addition of a fluorescent oxygen emission line at $\sim$0.53 keV improves the 
fits significantly. In accordance with earlier reports, we interpret the 
coronal component as emission from the planetary disk, produced by the 
scattering of solar X-rays in Saturn's upper atmosphere, and the line as 
originating from the Saturnian rings. The strength of the disk X-ray emission 
is seen to decrease over the period 2002 -- 2005, following the decay of 
solar activity towards the current minimum in the solar cycle. By comparing 
the relative fluxes of the disk X-ray emission and the oxygen line, we suggest 
that the line strength does not vary over the years in the same 
fashion as the disk flux. We consider possible alternatives for the origin of 
the line. The connection between solar activity and the strength of Saturn's 
disk X-ray emission is investigated and compared with that of Jupiter. 
We also discuss the apparent lack of X-ray aurorae on Saturn; by 
comparing the planet's parameters relevant to aurora production with those 
of Jupiter we conclude that Saturnian X-ray aurorae are likely to have gone 
undetected because they are below the sensitivity threshold of current 
Earth-bound observatories. A similar comparison for Uranus and Neptune leads
to the same disappointing conclusion, which is likely to hold true also with 
the planned next generation International X-ray Observatory. The next step
in advancing this research can only be realised with in-situ X-ray 
observations at the planets.}
   {}


   \maketitle
%

\section{Introduction}

In recent years X-ray observations have provided a novel way of furthering 
the study of planets in our solar system and our understanding of 
the physical processes taking place on them and in their environments (for 
a review see Bhardwaj et al. \cite{Bhardwaj_07} and references therein). 
For planets with a significant magnetic field, namely Jupiter and the Earth,
the X-ray emission has been found to separate essentially into two components:
a high-latitude auroral one, and a low-latitude disk component, which has 
different spectral properties, relating it to solar X-rays (Maurellis
 et al. \cite{Mau_00}, Gladstone et al.
\cite{Glad_02}, Branduardi-Raymont et al. \cite{BR_04}, Bhardwaj et al. 
\cite{Bhardwaj_solcon}, Elsner et al. \cite{El_05}, Branduardi-Raymont et al.
2007a,b, Bhardwaj et al. \cite{Bhardwaj_07}). Naively we would expect to 
observe a similar dichotomy on Saturn, given the strength of its magnetic 
field and its fast rotation (as for Jupiter) and its powerful UV aurorae
(G\'{e}rard et al. \cite{Gerard_04}, G\'{e}rard et al. \cite{Gerard_05}), but
this is not the case. While the characteristics of Jupiter's UV aurorae are 
dictated by the rotation of plasma internal to its magnetosphere (Bhardwaj
and Gladstone \cite{BG_00}, Clarke et al. \cite{Clarke_04}), those of Earth 
are solar wind driven (Clarke et al. \cite{Clarke_05}); Saturn's UV aurorae 
appear to be at variance with both these scenarios, rather than being 
intermediate between the two, as originally expected (Bhardwaj
and Gladstone \cite{BG_00}).
In Saturn's case the UV aurora responds strongly to the solar wind dynamic 
pressure (Crary et al. \cite{Crary_05}), whose compression of the magnetosphere
may induce reconnection leading to auroral brightenings (Cowley et al. 
\cite{Cowley_05}); compared to Earth, though, where aurora brightenings are 
very rapid ($\sim$ tens of minutes), those on Saturn can last for days (Clarke 
et al. \cite{Clarke_05}). Moreover, UV auroral activity has also been observed
on Saturn during a period (Oct. - Nov. 2005) of quiet solar wind conditions,
suggesting an intrinsically dynamic magnetosphere with injections of plasma
occasionally taking place in the night or dawn sectors (G\'{e}rard et al. 
\cite{Gerard_06}). Recent work by Clarke et al. (\cite{Clarke_09}) has added to
this picture: from an extensive campaign of observations by the Hubble Space 
Telescope (HST), coupled with planetary spacecraft and solar wind data, they 
find evidence for a direct relationship between Saturn's auroral activity and 
solar wind conditions, while the correlation is not so strong at Jupiter.

%
%
\begin{table*}
\caption{{\it XMM-Newton} and {\it Chandra} observations of Saturn}        
\label{table:1}      
\centering          
\begin{tabular}{c c c c c c l }   
\hline\hline       
 & Observation & Observation & Apparent & Heliocentric & Geocentric & 
  \\ 
Satellite & mid-epoch & duration & diameter & distance & distance & References
  \\
 &  &  (ks)  &  &  (AU)  &  (AU) &  \\
\hline \\                   
 {\it XMM-Newton} (1) & 2002 Oct. 01, 14:00 UT & 21 & 18.8'' & 9.0 & 8.8 & 
Ness et al. \cite{Ness_x}, this paper \\  
 {\it Chandra} (1) & 2003 Apr. 14, 18:00 UT & 66 & 17.5'' & 9.0 & 9.5 & 
Ness et al. \cite{Ness_c} \\
 {\it Chandra} (2) & 2004 Jan. 20, 05:30 UT & 37 & 20.5'' & 9.0 & 8.1 & 
Bhardwaj et al. \cite{Bhardwaj_05a}, \cite{Bhardwaj_05b}, \\
  &  &    &  &    &   &  this paper \\
 {\it Chandra} (3) & 2004 Jan. 26, 20:00 UT & 36 & 20.4'' & 9.0 & 8.2 & 
Bhardwaj et al. \cite{Bhardwaj_05a}, \cite{Bhardwaj_05b} \\ 
 {\it XMM-Newton} (2) & 2005 Apr. 22, 10:00 UT & 84 & 18.1'' & 9.1 & 
9.2 & This paper  \\
 {\it XMM-Newton} (3) & 2005 Oct. 28, 21:00 UT & 74 & 18.2'' & 9.1 & 
9.1 & This paper  \\
\\
\hline              
\end{tabular}
\end{table*}

Early observations of Saturn with {\it XMM-Newton} and {\it Chandra} (Ness et 
al. \cite{Ness_x}, Ness et al. \cite{Ness_c}) unambiguously detected X-ray 
emission, found it to be concentrated in the equatorial regions of the planet, 
and indicated that any X-ray auroral component, if present at all, is not 
easily detectable. More recent {\it Chandra} observations have provided us with
a remarkable view of the X-ray morphology of the Saturnian system, and of its 
temporal behaviour in the X-rays: both the planet's disk and the rings have 
been separately detected, and strong variability, closely following the 
behaviour of the solar X-ray flux, has been discovered in the disk X-ray 
emission (Bhardwaj et al. \cite{Bhardwaj_05a}, Bhardwaj et al. 
\cite{Bhardwaj_05b}). This suggests that its origin may lie in the scattering 
of solar X-rays in the upper layers of Saturn's atmosphere, as is the case 
for Jupiter's disk emission (Bhardwaj et al. \cite{Bhardwaj_solcon}, Bhardwaj
et al. \cite{Betal_06}, Cravens et al. \cite{Cravens_06}). Unlike 
Jupiter, though, no difference is found in the spectral characteristics 
of Saturn's disk and polar 
emissions, thus no evidence for X-rays of auroral origin is provided by the 
{\it Chandra} data. In fact, Bhardwaj et al. (\cite{Bhardwaj_05a}) 
remark that the X-ray emission from the polar regions appears to be 
anticorrelated with the strength of the FUV aurora, measured by the HST 
Imaging Spectrograph simultaneously with the {\it Chandra} 
observations. Moreover, only relatively little detail has been published 
so far about Saturn's X-ray spectral characteristics: an optically thin, 
coronal model is a good match for the disk emission (Ness et al. \cite{Ness_x},
Ness et al. \cite{Ness_c}, Bhardwaj \cite{Bhardwaj_06}), while the rings 
spectrum is practically made up of a single O K$\alpha$ emission line, which 
has been attributed largely to scattering of solar X-rays on the icy water 
ring material (Bhardwaj et al. \cite{Bhardwaj_05b}). However, solar 
fluorescence alone could not explain the ring X-ray brightness completely. 
To follow up on these issues, 
we made more extensive {\it XMM-Newton} observations of Saturn in 2005: 
this was a particularly appropriate time to take a deeper look at the planet, 
while a campaign of HST observations was taking place, 
and the {\it Cassini} spacecraft (which entered orbit around Saturn in July 
2004) was making coordinated in-situ measurements of a variety of physical 
parameters in its environment. 

In this paper we report the results from our 2005 observations, 
together with a re-analysis of the earlier (2002) data extracted from the 
{\it XMM-Newton} archive (sec.s 2 and 3). We also model the {\it Chandra} 
X-ray spectra from the 2003 and 2004 observations (which have already been
described by Ness et al. \cite{Ness_c}, Bhardwaj et al. \cite{Bhardwaj_05a} 
and Bhardwaj et al. \cite{Bhardwaj_05b} $-$ see sec. 3). We discuss our 
findings in sec. 4, also making comparisons with Jupiter and Earth, as well 
as Uranus and Neptune, and draw our conclusions in sec. 5. 

\section{{\it XMM-Newton} observations of Saturn}   

In Table~1 we list in time order all the X-ray observations of Saturn which 
have been carried out so far with {\it XMM-Newton} and {\it Chandra}, 
identifying them by sequential numbers; those with {\it Chandra}
were all performed with the Advanced CCD Imaging Spectrometer (ACIS), so that 
spectral information was available. Table~1 includes the mid-epochs and 
durations of the observations, as well as the apparent size of the planet's 
disk, its distance from the Sun and the Earth at those times, and
relevant references. During our 2005 {\it XMM-Newton} observations
the CCD European Photon Imaging Cameras (EPIC-pn, Str\"{u}der et al.
\cite{struder}, EPIC-MOS, Turner et al. \cite{turner}) were operated in Full 
Frame mode with the thick filter at all times, to avoid any optical leak 
contamination (see Ness et al. \cite{Ness_x} for a discussion of the optical 
light leak affecting thin and medium filters). In 2002 only one exposure (with 
the EPIC-pn camera) was carried out with the thick filter, so the analysis had 
to be restricted to it for this epoch. In order to make a direct and consistent
comparison of all {\it XMM-Newton} data, the analysis of the 2005 observations 
reported here is also limited to the EPIC-pn camera. We note, though, that the 
best fit of the combined EPIC-pn and MOS datasets for 2005 is consistent,
within the parameters errors, with that obtained using the EPIC-pn camera on
its own. During all {\it XMM-Newton} observations
the Reflection Grating Spectrometer (RGS, den Herder et al. \cite{denHerder}) 
was run in Spectroscopy mode, but the source brightness is too low to enable
high resolution spectroscopy to be carried out. The Optical Monitor 
telescope (Mason et al. \cite{Mason}) had its 
filter wheel kept in the BLOCKED position, because Saturn's optical brightness 
exceeds the safe limit for the instrument (thus no OM data were collected).

At all 2002 and 2005 observing epochs Saturns's path on the sky was
essentially in RA, and very slow (8''/hr max). Each one of the longer,
2005 observations was broken into two intervals, with a small pointing
trim in between, to ensure the planet would be imaged on the same CCD chip
at all times. The two intervals were combined during the data processing and 
the subsequent analysis.

%
   \begin{figure}
   \centering
   \includegraphics[width=5.5cm,angle=0]{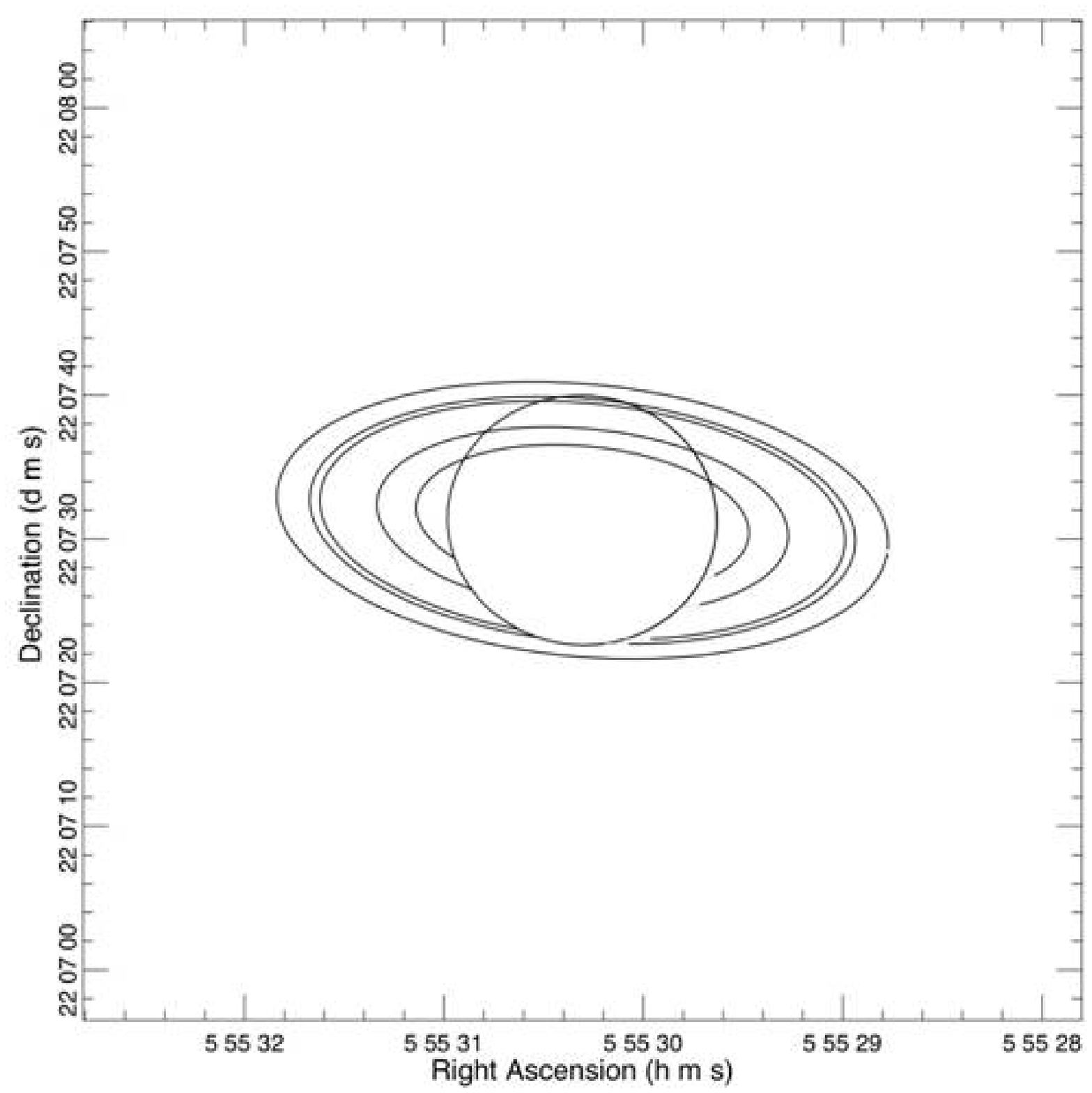}

    \includegraphics[width=5.5cm,angle=0]{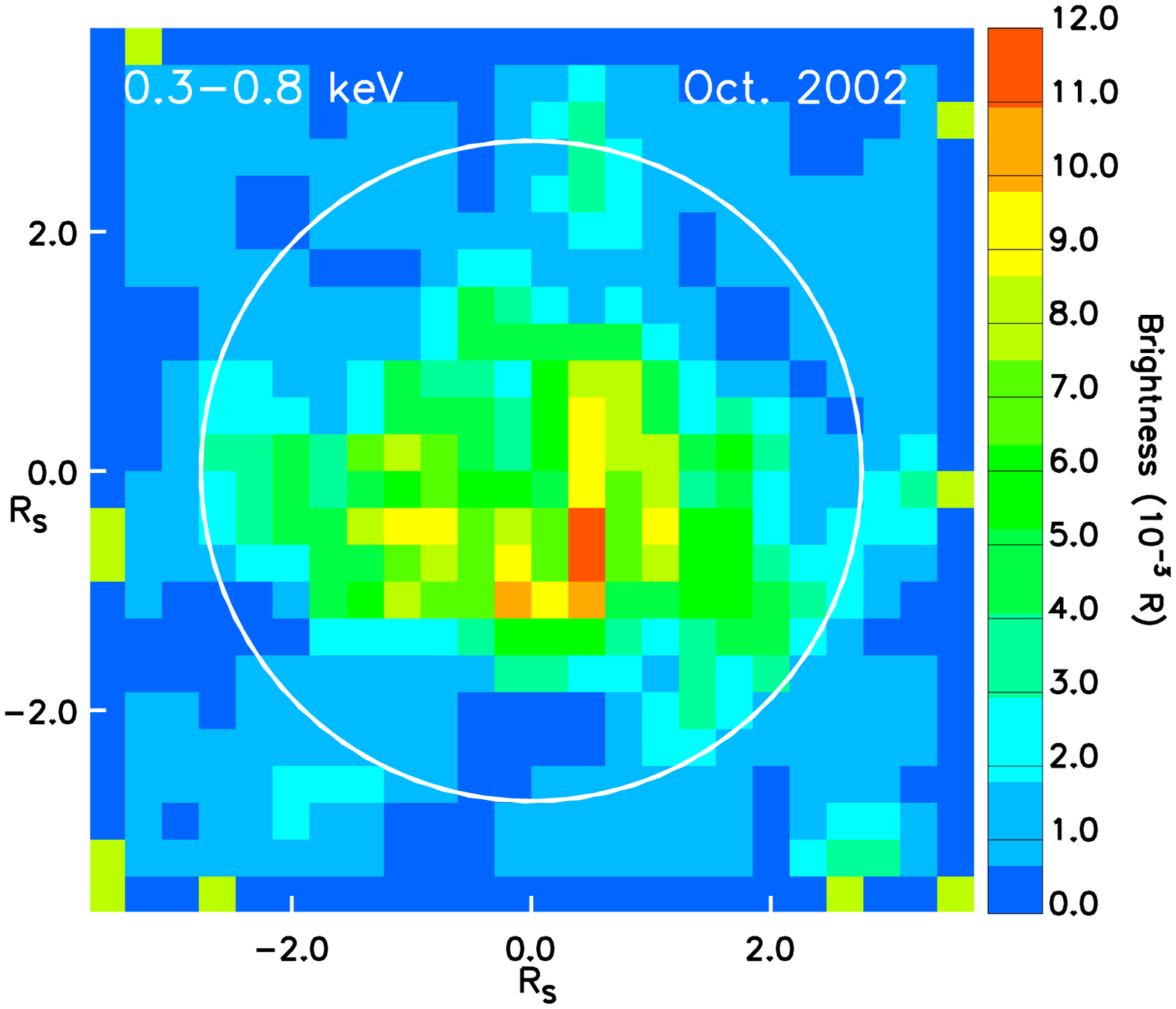}

    \includegraphics[width=5.5cm,angle=0]{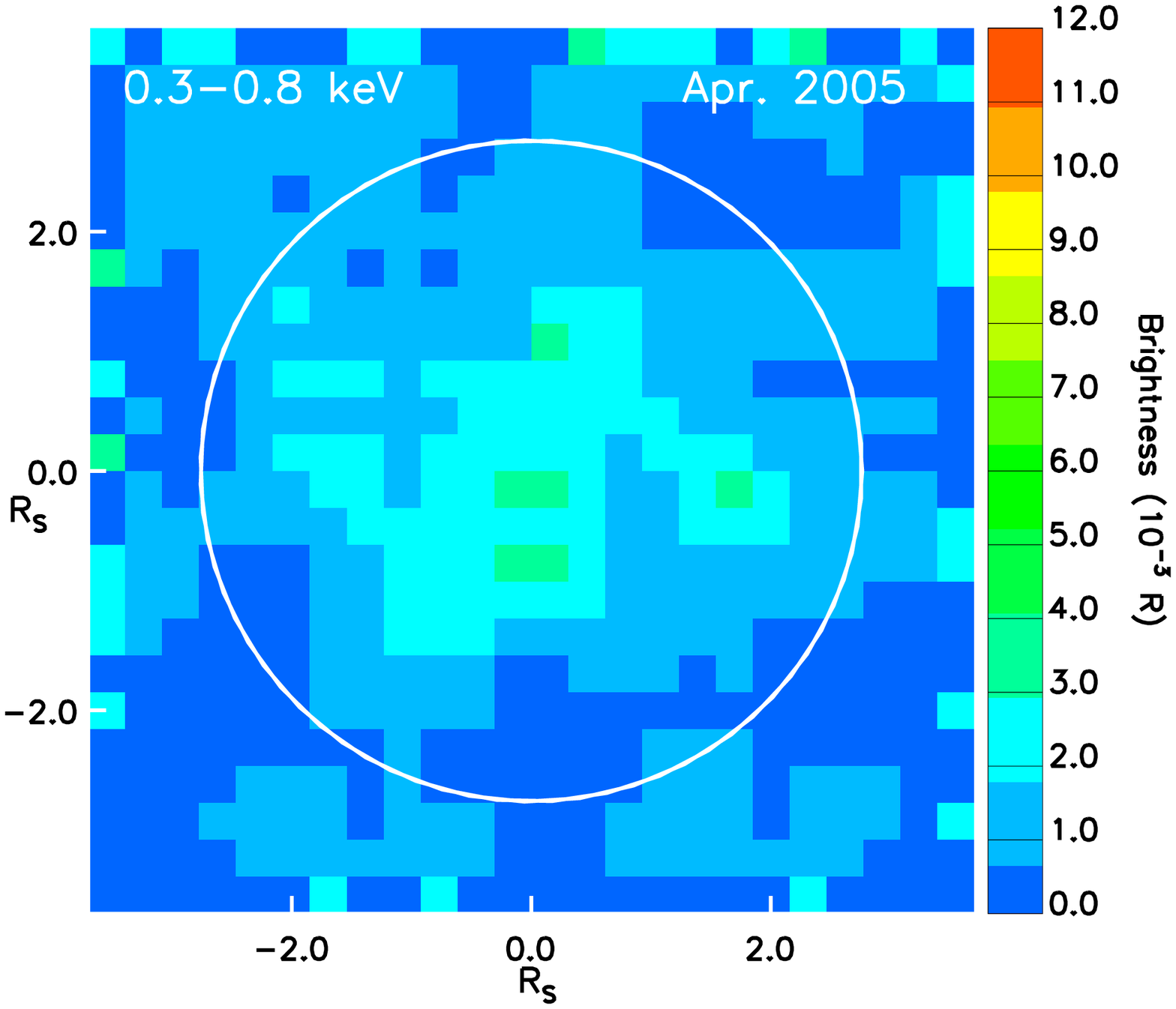}

    \includegraphics[width=5.5cm,angle=0]{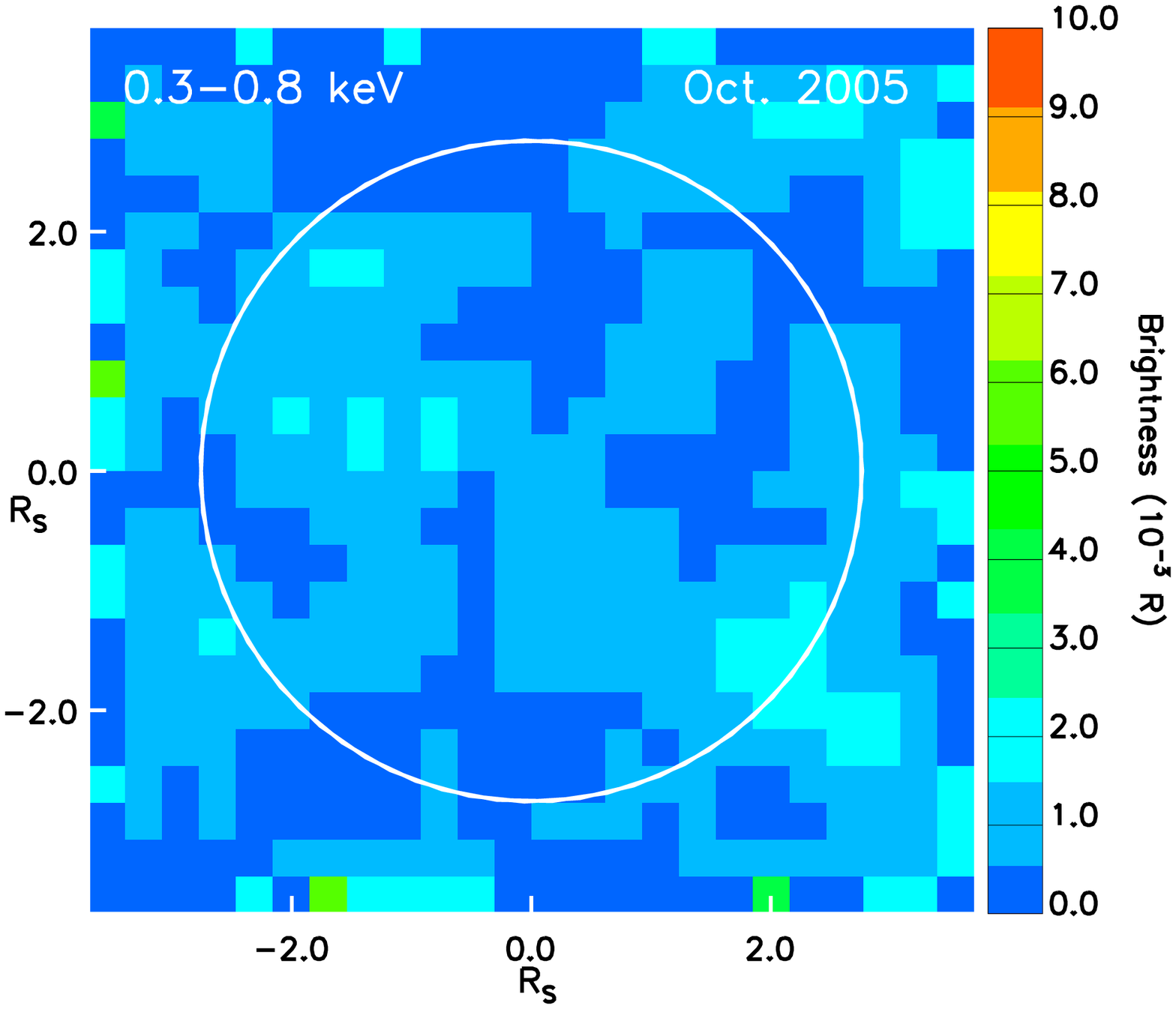} 
     \caption{The black and white diagram in the top panel shows the 
              appearence of Saturn and its rings (outer ellipse major axis =
              44'') at the epoch of the 
              {\it XMM-Newton} observations in Oct. 2002 (the view in 2005
              was very similar). In the three colour panels are 
              {\it XMM-Newton} EPIC-pn images of Saturn in the 0.3 -- 0.8 keV 
              band for Oct. 2002 (top), Apr.2005 (middle) and Oct. 2005 
              (bottom). Overplotted in white are
              26'' radius circles centred on the planet, delimiting the area
              used for spectral extraction. The size of the 
              diagram and of the images is 70'' x 70'', and the 
              EPIC-pn image pixels are of 2.9'' side (North is at the top and 
              East to the left). It is clear that the planet was X-ray 
              brightest in 2002 and weakest, in fact undetected, in Oct. 2005.
              }
         \label{Fig1}
   \end{figure}

The two {\it XMM-Newton} 2005 observations, as well as the archival one 
from 2002, were processed and analysed in a uniform fashion under SAS (Science
Analysis Software) v. 7.1; the 2002 observation (Ness et al. \cite{Ness_x}) 
had never been processed in this now standard way, because of the lack, at the 
time, of dedicated software to re-register the {\it XMM-Newton} events in 
the rest frame of a moving target. After this was done for the three Saturn
datasets, images at the location of the planet were constructed with the
data from the EPIC cameras. Fig. 1 shows the EPIC-pn images of Saturn 
in the energy range 0.3 -- 0.8 keV for the {\it XMM-Newton} observations 
1 (top of the colour panels), 2 (middle) and 3 (bottom; the planet is
undetected on this occasion). Overposed to each image, at the nominal 
location of the planet's centre, is a white circle of 26'' radius. At the 
top of Fig. 1 is a diagram showing the appearence of Saturn and its rings 
in Oct. 2002 (the view, with the South Pole unobstructed, was very similar 
in Apr. and Oct. 2005). 

From visual comparison of the three panels in Fig. 1 it is clear that the 
planet was X-ray brightest in 2002, dimmer in Apr. 2005 and weakest, in fact
undetected, in Oct. 2005 (see sec. 3 and Table 2 for X-ray upper limits 
at this epoch). 

Despite the relatively broad {\it XMM-Newton} Point Spread Function (PSF, of 
15'' Half Energy Width) compared with the planet's image size ($\sim$20''
diameter), the morphology of the X-ray emission at both 2002 and 
Apr. 2005 epochs does not appear circularly symmetric, but is seen 
to stretch along the direction of the rings plane. In the 
following we will analyse the {\it XMM-Newton} data as encompassing both, 
disk and ring emissions. 

We also searched for evidence of X-ray emission from Saturn's satellite Titan 
in the {\it XMM-Newton} datasets, but none was found. Similarly no detection 
was obtained with the {\it Chandra} observations.

\section{Spectral fitting: {\it XMM-Newton} EPIC and {\it Chandra} ACIS data}

In principle Saturn occults the cosmic X-ray background originating
beyond it, so we could expect that background subtraction may not be 
required in the analysis of its spectra; however, the contribution of 
particle background (dominant at higher energies), and any `spill over' 
of the cosmic X-ray background onto the planet's disk due to the width of 
{\it XMM-Newton} PSF may become significant if we use a region to 
extract the X-ray events for the planet which is large compared with 
the size of its disk. This is our case, where we are using an extraction
circle of 26'' radius, in order to include as much as possible of the 
X-ray flux from the source (Saturn's disk diameter $\simeq$ 20''). For an
{\it XMM-Newton} point
source (which, for this purpose, Saturn can be approximated to), an 
extraction radius of 26'' corresponds to an encircled energy of 80\%, so all 
fluxes quoted in the following for the planet are corrected for this factor. 
Fig. 2 shows a comparison of Saturn's 2002 and Apr. 2005 EPIC-pn spectra 
(black crosses) extracted from the regions delimited by white circles in 
Fig. 1, with those (red crosses) from background regions (devoid of sources) 
of the same shape and size, from the same EPIC-pn CCD. The plots indicate 
that the emission from the planet lies above the background in the band 0.3 
-- 2 keV, and that it practically disappears under it at higher energies.
The {\it XMM-Newton} spectral fitting results presented below refer to the 
Oct. 2002 and Apr. 2005 observations (the planet was undetected in Oct. 2005).

In the case of the {\it Chandra} data, unlike the analysis carried out by 
Bhardwaj et al. (2005a,b),
we extracted the full planet data, i.e. the combined emission from the 
disk, the polar cap and the rings, for the purpose of comparing
with {\it XMM-Newton} which cannot resolve these morphological details.  
As already pointed out by Bhardwaj et al. 
(\cite{Bhardwaj_05a}), the spectrum of Saturn's polar region is very similar
to that of the planet's disk. Here we present results from the fits made to 
the Apr. 2003, the 20 and the 26 Jan. 2004 {\it Chandra} data. The 2003
observation was made up of two exposures: we show the best fits for the two 
datasets combined, since individually they give very similar results. The 
source was weaker by a factor of three on 26 Jan. 2004 than a week earlier
(Bhardwaj et al. \cite{Bhardwaj_05a}), when Saturn brightened in the X-rays 
in direct response to a powerful X-ray flare taking place on the solar 
hemisphere visible from the planet.
 

%
   \begin{figure}
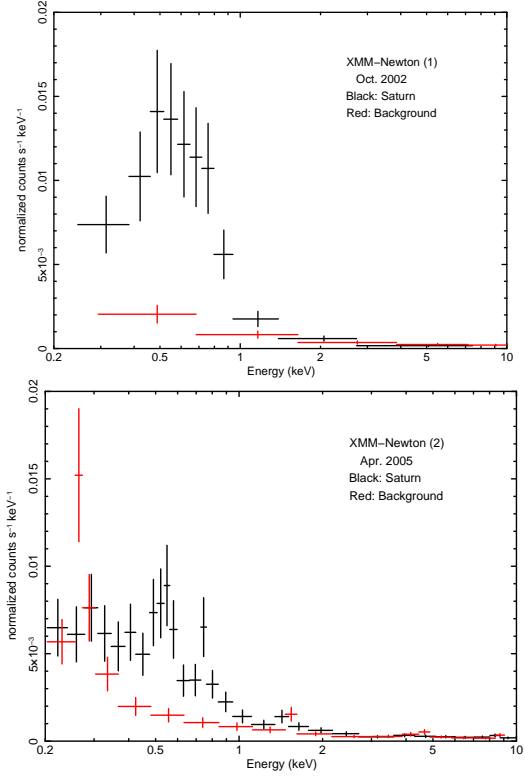

   \centering
   \includegraphics[width=5cm,angle=-90]{sat_back.cps}
\\
    \includegraphics[width=5.2cm,angle=-90]{sat_back_apr05.cps}
       \caption{{\it XMM-Newton} EPIC-pn spectra of Saturn (black crosses,
                 before any background subtraction) extracted 
                 from the regions marked by white circles in Fig. 1; in red
                 are the spectra of the background, extracted from regions 
                 devoid of sources, of the same circular shape and size as for
                 Saturn, and from the same EPIC-pn CCD (top: Oct. 2002; 
                 bottom: Apr. 2005).
              }
         \label{Fig2}
   \end{figure}
%

Before fitting, the spectra were binned so as to include at least 15 counts 
per energy channel (10 counts for the low-flux {\it Chandra} observation of 
26 Jan. 2004), to validate the applicability of $\chi^2$ testing 
for parameter estimation (Cash \cite{cash}). We used XSPEC v. 12.4.0 for
the fitting and modeled the spectra with an optically thin 
{\tt mekal} component (to represent the scattered solar X-ray contribution) 
and a Gaussian line at 0.53 keV to take into account the O K$\alpha$ 
line emission from the rings. The {\tt mekal} temperature and normalisation, 
and the energy and normalisation of the emission line were left free
to vary in the fit. Elemental abundances for the coronal model were set at the 
solar level; the width of the emission line was kept fixed, and taken to 
be due only to instrumental broadening. This model combination 
provides good fits to the EPIC-pn spectra from the {\it XMM-Newton} 
observations in 2002 and Apr. 2005, and also to the {\it Chandra} ACIS 
spectra for Saturn and its rings combined, from the 2003 and 2004 observations.
Fig. 3 shows the data (black crosses) and the best fits (red histograms) for 
all the {\it XMM-Newton} and {\it Chandra} datasets in time sequence (from 
top to bottom).

   \begin{figure}
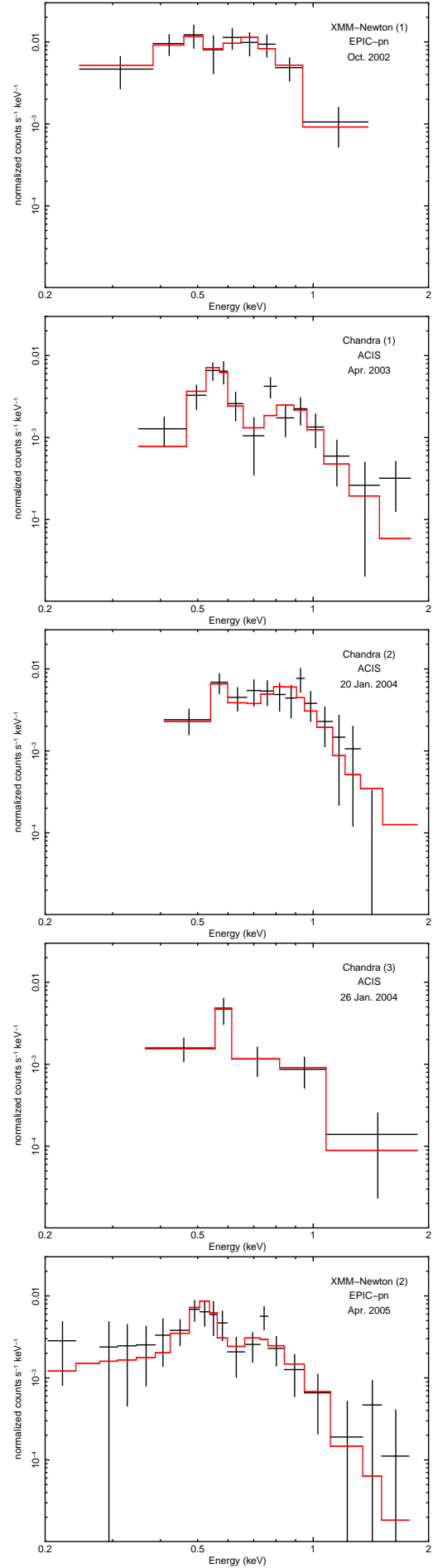

   \centering
    \includegraphics[width=5cm,angle=-90]{XMM_1_line_fit.cps}
\\
    \includegraphics[width=5cm,angle=-90]{chandra03_ron_total_best_fit.cps}
\\
    \includegraphics[width=5cm,angle=-90]{20Jan_ron_best_fit.cps} 
\\ 
    \includegraphics[width=5cm,angle=-90]{26Jan_ron_best_fit.cps} 
\\ 
    \includegraphics[width=5cm,angle=-90]{XMM_2_pn_line_fit.cps}
\\
       \caption{{\it XMM-Newton} EPIC-pn and {\it Chandra} ACIS spectra of 
               Saturn, in time sequence from top to bottom. The data are 
               shown as black crosses, the best fits as red histograms.
            }
         \label{Fig3}
   \end{figure}
%

Table 2 lists the total numbers of source counts in the spectra, the best fit 
parameters and their 90\% confidence errors,
as well as energy fluxes for the two spectral components separately, for all 
the spectra from the two observatories. The fluxes were obtained from the best 
fit models by setting to zero the normalisation of one or the other of the 
components and have been converted 
to emitted powers (also listed in Table 2) by multiplying by 4$\pi${\it d}$^2$,
where {\it d} is the geocentric distance of Saturn at each epoch. 

We have checked whether the presence of the oxygen emission line in the spectra
is statistically significant by setting its normalisation to 0 and re-fitting 
the spectra. The resulting best fit $\chi^2$ values are listed in the last 
column of Table 2 and demonstrate that ignoring the presence of the line 
worsens the fits significantly.

Both Apr. 2003 and Apr. 2005 spectra in Fig. 3 show one bin deviating
in excess of the best fit at $\sim$0.75 keV (16.5 \AA); in both cases this 
feature can be accounted for by a narrow emission line at the instrumental 
resolution. Interestingly, Fe XVII lines at 15 and 17 \AA\ are characteristic
of the solar coronal spectrum; however, the statistical quality of the data 
(and in particular the hard-to-constrain strength of the line) does
not warrant exploring this further.

The {\it XMM-Newton} (3) observation (Oct. 2005) shows no evidence of X-ray 
emission from the planet (see bottom panel of Fig. 1), and we extract a 
3$\sigma$ upper limit of 44 counts in the EPIC-pn camera over the full duration
(74 ks) of the dataset. Adopting the spectral parameters that best fit the 
Apr. 2005 {\it XMM-Newton} (2) EPIC-pn observation, assuming the same relative
strengths for the disk and line emissions, and normalising to the shorter
exposure and the slightly smaller geocentric distance, we obtain an upper limit
of 19 MW to the disk emitted power in Oct. 2005, down by more than a factor 
of 10 from the level observed three years earlier during the {\it XMM-Newton} 
(1) observation (Table 2). 

\section{Discussion}

\subsection{Separating Saturn's disk and rings X-ray emission}

Our spectral fitting of Saturn's {\it XMM-Newton} and {\it Chandra} data,
gathered over the period Oct. 2002 $-$ Oct. 2005, shows that the X-ray
emission is well described by a combination of two components, an optically 
thin coronal model ({\tt mekal}) and an O K$\alpha$ emission line; in 
accordance with the earlier results of Ness et al. (2004a,b) and Bhardwaj 
et al. (\cite{Bhardwaj_05a}), we interpret the 
{\tt mekal} component in Saturn's X-ray spectrum as representing its disk 
emission, which varies over the years 2002 -- 2005 according to the 
diminishing strength of the solar X-ray output. The average temperature of 
the coronal emission is 0.5 keV, which is similar to that measured 
(0.4 $-$ 0.5 keV) on three occasions in 2003 for Jupiter's disk X-rays 
(Branduardi-Raymont et al. \cite{BR_07a}); 
this strengthens the idea that in both cases the emission originates from the 
scattering of solar X-rays in the planets' upper atmospheres (cf. Bhardwaj et
al. 2005a,c, Bhardwaj 2006, Cravens et al. \cite{Cravens_06}).

The presence of an additional spectral component, in the form of an 
oxygen emission line, had already been suggested by the early fits of the 
2002 and 2003 {\it XMM-Newton} and {\it Chandra} data by Ness et al. (2004a,b);
this line was later associated with the planet's rings by Bhardwaj et al. 
(\cite{Bhardwaj_05b}) using the {\it Chandra} data, by spatially selecting 
only events coincident with the Saturnian rings. 
We have taken the different approach of attempting to separate spectrally 
the two emission components through a uniform analysis of the spectra of 
the whole planet + rings system from both observatories. This solves 
the problem of the spatial resolution of {\it XMM-Newton} being too limited
to allow us to separate spatially the planet's disk from the rings; then we
can study the variability of both emission components over the solar cycle
by examining {\it XMM-Newton} and {\it Chandra} spectral fitting results,
under the assumption that all the line emission originates in the rings.

   \begin{figure}
   \centering
    \includegraphics[width=9.2cm,angle=0]{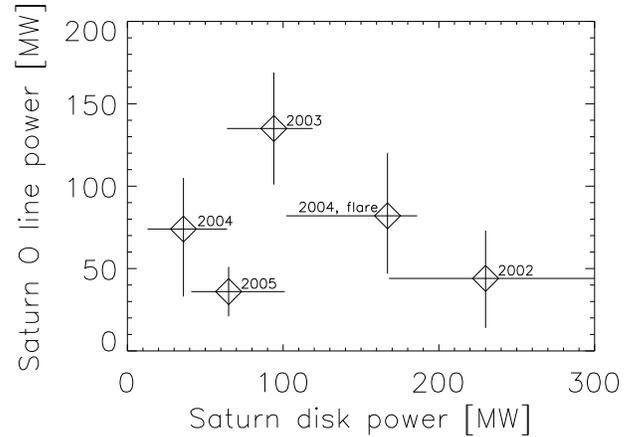}
       \caption{Saturn's power emitted in the oxygen line plotted 
                against the 0.2 $-$ 2.0 keV power from the planet's disk,
                for all {\it XMM-Newton} and {\it Chandra} observations.
            }
         \label{Fig4}
   \end{figure}
%

As a consistency check, we have compared the parameters of our best fits and 
the values of emitted 
power derived from them (Table 2) with those previously published for the 
2002 {\it XMM-Newton} observation and the 2003 and 2004 {\it Chandra} datasets.
In the first case, Ness et al. (\cite{Ness_x}) use the {\tt apec} model in 
XSPEC to describe the scattered solar spectrum, and combine it with a narrow 
oxygen line: their best fit temperature of 0.33 keV is consistent with our 
value of 0.28 keV (for a {\tt mekal} model) within the errors; they derive 
{\tt apec} and line fluxes of 1.36 and 0.22 $\times$ 10$^{\rm -14}$ erg 
cm$^{\rm -2}$ s$^{\rm -1}$ respectively, which are 30\% and 10\%  higher than 
our {\tt mekal} and line fluxes respectively, but again consistent within the 
spectral parameters errors.
For the {\it Chandra} 2003 observation, Ness et al. (\cite{Ness_c}) fit only
the emission from the circular region coinciding with the planet's disk 
(30\% of which is occulted by the rings), 
deriving a best fit {\tt mekal} temperature of 0.39 keV, with a 0.1 $-$ 2 keV 
flux of 0.6 $\times$ 10$^{\rm -14}$ erg cm$^{\rm -2}$ s$^{\rm -1}$, 
and a line flux of 0.13 $\times$ 10$^{\rm -14}$ erg cm$^{\rm -2}$ s$^{\rm -1}$.
In this case our results are somewhat different (see Table 2) because we 
include the full rings emission in the fit (thus our oxygen line is stronger).
We can make a more direct comparison with the results obtained by Bhardwaj 
et al. (\cite{Bhardwaj_05b}) on the rings line emission observed by 
{\it Chandra}. These authors quote a line flux of 0.5 $\times$ 10$^{\rm -14}$
erg cm$^{\rm -2}$ s$^{\rm -1}$ for the 20 Jan. 2004 observation which is 
in very good agreement with our value (0.48 $\times$ 10$^{\rm -14}$ erg 
cm$^{\rm -2}$ s$^{\rm -1}$). Note that their flux is derived by fitting the 
rings spectrum on its own in XSPEC, while here we obtain the rings flux by
extracting and modelling
the combined disk and rings spectrum (in order to compare with the 
un-separated spectra from {\it XMM-Newton}). The consistency of the two 
results gives confidence in our approach. However, the line powers which 
we derive from the 2003 and 26 Jan. 2004 {\it Chandra} data are factors 
of $\sim$ 2 and 1.7 above what Bhardwaj et al. (\cite{Bhardwaj_05b}) quote
respectively. A possible explanation of the discrepancy may be the presence 
of oxygen line emission from the rings area occulting the disk, which was 
excluded by Bhardwaj et al. but is included in our analysis. In fact, Ness 
et al. (\cite{Ness_c}) find that the inclusion of a small amount of line 
emission does improve their fit of the {\it Chandra} 2003 full disk region 
(including the part covered by the rings). If this is the case, the additional
line contribution from this rings area may become more relevant when the disk 
flux is lower, as indeed can be seen in the {\it Chandra} observations (1) 
and (3) (Fig. 3).
 
Turning to the power emitted by the disk, inspection of Bhardwaj et al. 
(\cite{Bhardwaj_05a}) Fig. 4 indicates 0.2 $-$ 2.0 keV powers of 
$\sim$380, 170, 250 and 90 MW in 2002, 2003 and on 20 and 26 Jan. 2004 
respectively; these are a factor between 1.5 and 2.4 larger than those 
we derive and list in Table 2. The different analysis approach may be 
responsible for this discrepancy. However, since we are interested in 
studying trends over time, our uniform approach to the whole dataset 
ought to give more comparable results.

\begin{landscape}
\headsep = 190pt
\begin{table*}
\centering
\vspace{0.5truecm}
\caption{{\it XMM-Newton} and {\it Chandra} spectral fitting results (errors 
are at 90\% confidence)}         
\label{table:2}      
\begin{tabular}{c c c c c c c c c c c c c c}   
\hline\hline      
 Year & Satellite & Source & 
 {\tt mekal} & {\tt mekal} & {\tt mekal} & {\tt mekal} & Gauss & Gauss & Gauss
& Gauss & $\chi^2$ & $\chi^2$ & GOES-10 \\ 
of & (observation & net &  kT$^a$ & norm$^b$ & flux$^c$ & 0.2 $-$ 2.0 
keV emitted & energy & norm$^e$ & flux$^f$ & emitted &/d.o.f.$^g$ &/d.o.f.$^g$ 
 & 1 -- 8 \AA \\
 obs. & number) & counts & & & & power$^d$ (MW) & (keV) & & & 
power$^d$ (MW) & & no line & flux$^i$ 
\\
\hline
\\          

\vspace{1 mm}
          
2002 & {\it XMM-Newton} (1) & 112 &0.28$^{\rm +0.06}_{\rm -0.04}$ 
 & 4.3$^{\rm +1.2}_{\rm -1.1}$ & 1.04$^{\rm +0.36}_{\rm -0.28}$ & 
 230$^{\rm +80}_{\rm -62}$ & 0.48$^{\rm +0.04}_{\rm -0.05}$ 
 & 2.6$^{\rm +1.7}_{\rm -1.8}$ &0.20$^{\rm +0.13}_{\rm -0.14}$ & 
 44$^{\rm +29}_{\rm -30}$ & 0.9/5 & 7.0/7 & 1.13 \\

\vspace{1 mm}

2003 & {\it Chandra} (1) & 137 & 0.65$\pm${0.19}  
 &1.3$\pm${0.4} & 0.37$^{\rm +0.10}_{\rm -0.12}$ & 
 94$^{\rm +25}_{\rm -30}$ & 0.55$^{\rm +0.02}_{\rm -0.01}$
 & 5.9$\pm${1.5} & 0.53$\pm${0.13} & 135$\pm${34} & 8.6/9 & 32.6/11 & 2.12 \\

\vspace{1 mm}

2004 & {\it Chandra} (2) & 112 & 0.59$^{\rm +0.15}_{\rm -0.28}$  & 
 3.1$^{\rm +0.7}_{\rm -0.8}$ & 0.90$^{\rm +0.10}_{\rm -0.35}$ & 
 167$^{\rm +19}_{\rm -65}$ & 0.57$\pm${0.03}
 & 4.9$^{\rm +2.3}_{\rm -2.1}$ & 0.44$^{\rm +0.21}_{\rm -0.19}$ & 
 82$^{\rm +38}_{\rm -35}$ & 12.2/10 & 16.7/12 & 3.23 \\

\vspace{1 mm}

2004 & {\it Chandra} (3) & 42 & 0.61$^{\rm +0.84}_{\rm -0.36}$  
 & 0.7$^{\rm +0.8}_{\rm -0.4}$ & 0.19$^{\rm +0.15}_{\rm -0.12}$ & 
 36$^{\rm +28}_{\rm -23}$ & 0.56$^{\rm +0.05}_{\rm -0.03}$
 & 4.3$^{\rm +1.8}_{\rm -2.4}$ & 0.39$^{\rm +0.16}_{\rm -0.22}$ & 
 74$^{\rm +31}_{\rm -41}$ & 0.22/1 & 5.2/3 & 0.16 \\

\vspace{1 mm}

2005 & {\it XMM-Newton} (2) & 146 & 0.33$^{\rm +0.12}_{\rm -0.08}$ &
 1.1$\pm${0.4} & 0.27$^{\rm +0.15}_{\rm -0.10}$ & 
 65$^{\rm +36}_{\rm -24}$ & 0.52$\pm${0.02} & 1.7$\pm${0.7} & 
 0.15$\pm${0.06} & 36$\pm${15} &
 17.4/16 & 33.3/18 & 0.22 \\

2005 & {\it XMM-Newton} (3) & $<$ 44$^h$ & 0.33 & $<$ 0.3$^h$ 
& $<$ 0.08$^h$ & $<$ 19$^h$ & 0.52 & $<$ 0.51$^h$ & $<$ 0.05$^h$ & $<$ 12$^h$ 
& & & 0.004 \\
\\
\hline             
\end{tabular}
\begin{list}{}{}
\item[$^{a}$] {\tt mekal} temperature in keV\\
\item[$^{b}$] {\tt mekal} normalisation at 1 keV in units of 
  10$^{\rm -6}$ ph cm$^{\rm -2}$ s$^{\rm -1}$ keV$^{\rm -1}$\\
\item[$^{c}$] Total 0.2 -- 2.0 keV energy flux in the {\tt mekal} component 
  in units of 10$^{\rm -14}$ erg cm$^{\rm -2}$ s$^{\rm -1}$; the errors
  on the fluxes were obtained by varying the {\tt mekal} temperatures
  and normalisations between their 90\% confidence limits  \\
\item[$^{d}$] Powers and errors calculated by multiplying the values 
  in the previous column by 4$\pi${\it d}$^2$, where {\it d} is the geocentric 
  distance of Saturn at the given epoch.
\item[$^{e}$] Gaussian line normalisation in units of 
  10$^{\rm -6}$ ph cm$^{\rm -2}$ s$^{\rm -1}$ \\
\item[$^{f}$] Total energy flux in the line in units of
  10$^{\rm -14}$ erg cm$^{\rm -2}$ s$^{\rm -1}$\\
\item[$^{g}$] $\chi^2$ value and degrees of freedom\\
\item[$^{h}$] 3$\sigma$ upper limit
\item[$^{i}$] Solar X-ray flux (1.5 -- 12.4 keV) in units of 10$^{\rm -6}$
  Watt m$^{\rm -2}$, corrected for solar rotation (see text)
\end{list}
\end{table*}

\begin{table*}
\centering
\caption{Planetary parameters relevant to auroral studies}             
\label{table:3}      
\begin{tabular}{c c c c c c c c c c}   
\hline\hline       
Planet & Average solar & Rotation & Equatorial & Surface & Magnetic & 
FUV aurora     & X-ray aurora & Disk X-ray & Magnetospheric \\ 
       & distance  & period & radius & magnetic field & moment & 
average input & total average & average & ions max. \\
& (AU)$^a$ & (hrs)$^a$ & (Earth = 1)$^{a,b}$ & (G)$^a$ &(Earth = 1)$^a$ &power 
(GW)$^c$ & power (MW)$^d$ & power (MW) & density (cm$^{\rm -3}$)$^a$ \\
\hline
\\                    
 Earth   & 1.0 & 23.9 & 1 & 0.31 & 1& 1 $-$ 100 &10 $-$ 30  & 40$^d$ & 
1 $-$ 4000  \\
 Jupiter & 5.2 & 9.9  & 11.2 & 4.28 & 20,000 & few $\times$ 10$^4$ & 400 $-$ 
1000  & 500 $-$ 2000$^d$  & $>$ 3000   \\
 Saturn  & 9.6 & 10.7 & 9.4 & 0.22 & 600  & 10$^2$ $-$ 10$^3$ & ? & $\leq$ 
230$^e$ & 100 ?   \\  
 Uranus  & 19.2 & 17.2 & 4.0 & 0.23 & 50 & $\lesssim$40$^f$ & ? & ? & 3 \\
 Neptune & 30.1 & 16.1 & 3.9 & 0.14 & 25 & $\lesssim$0.1$^f$ & ? & ? & 2 \\
\\
\hline             
\end{tabular}
\begin{list}{}{}
\item[$^{a}$] From {\it The New Solar System}, Ed.s J. K. Beatty, C. Collins 
Petersen \& A. Chaikin, 1999, Cambridge University Press
\item[$^{b}$] For the outer planets, radius at 1 bar atmospheric pressure
\item[$^{c}$] Clarke et al. (\cite{Clarke_05})
\item[$^{d}$] Bhardwaj et al. (\cite{Bhardwaj_07})
\item[$^{e}$] This paper
\item[$^{f}$] Bhardwaj and Gladstone (\cite{BG_00})
\end{list}
\end{table*}
\end{landscape}

Then we can use the whole sequence of {\it XMM-Newton} and {\it Chandra} 
observations to study how the fluxes in the two X-ray spectral components,
the emission from the planet's disk and the oxygen line (assumed to be 
confined to the rings), varied relative to each other over a period of 
three years. From Fig. 3 it is already clear that the line is more prominent 
when the disk flux is lower (Apr.
2003, 26 Jan. 2004 and Apr. 2005) and is less visible when the disk is
brighter (Oct. 2002 and 20 Jan. 2004). This suggests that the two
components vary over time in different ways: Fig. 4 shows the power
emitted in the oxygen line plotted against that from the disk (from 
Table 2). 
The trend of the disk emission to decrease with time (mirroring the decay
of the solar X-ray output as the solar cycle progresses towards a minimum),
originally identified by Bhardwaj et al. (\cite{Bhardwaj_05a}), 
is apparent from the temporal sequence of the data in Fig. 4. However,
despite the large errors, it is clear that there is no similar trend in the 
oxygen line; although both disk and line powers change by factors of $\sim$ 
6 and 4 respectively over the three years of observations, their 
variability appears to be un-correlated: in particular, the line strength 
is consistent with being the same in 2002 and Apr. 2005, while the disk 
has dimmed by a factor of 3.5. In addition, as reported by Bhardwaj et al. 
(\cite{Bhardwaj_05b}), the response of the disk to the solar flare on 20 Jan.
2004 is not clearly matched by a similar increase in the rings line power. 
Both disk and rings have disappeared by Oct. 2005, which may argue for some 
correlation between the two; however, it is not clear that the line, if 
present but weak, would produce enough counts to be detectable on its own. 

\subsection{The solar connection}

Fig. 5 shows Saturn's disk 0.2 -- 2.0 keV X-ray power from Table 2 (open 
squares) plotted against the solar 1 $-$ 8 $\AA$ (1.5 $-$ 12.4 keV) flux 
(also listed in Table 2) measured by the GOES-10 satellite at each 
observing epoch and averaged over the duration of the observations.
These fluxes have been corrected for the Sun's rotation except in the cases 
(Jan. 2004) when the Earth and Saturn were practically in opposition. 
We have neglected the travel time differences Sun-Earth and Sun-Saturn-Earth 
because these are very small with respect to the correction for solar rotation 
(and cancel out in opposition). As expected, there appears to be correlation 
between Saturn's disk X-ray power and the solar X-ray flux. For comparison,
the same parameters are also plotted for Jupiter's disk (filled circles),
which also has been shown to act as a mirror of solar X-rays (Maurellis et al.
\cite{Mau_00}, Bhardwaj et al. \cite{Bhardwaj_solcon}, Cravens et al. 
\cite{Cravens_06}). {\it XMM-Newton} and {\it Chandra} observations in 
Fig. 5 are identified by an X and a C, respectively. 

For completeness here is a brief summary of Jupiter's measurements used in 
Fig. 5. Waite 
et al. (\cite{Waite_97}) were the first to report an X-ray measurement for the 
low latitude regions of the planet from their {\it ROSAT} High Resolution 
Imager observation in July 1994. {\it Chandra} observed Jupiter with the
High Resolution Camera (HRC) in Dec. 2000, with both the HRC and ACIS in Feb. 
2003 and with ACIS in Feb. -- March 2007; this last time the observations were 
simultaneous with {\it XMM-Newton}, which had looked at the planet also 
in April and Nov. 2003. Bhardwaj et al. (\cite{Betal_06}) report 0.5 -- 1.5 keV
fluxes for the two 2003 ACIS observations, which we have converted to
0.2 -- 2.0 keV using their {\tt mekal} best fit parameters. They also 
mention that the HRC flux was about 50\% of that in Dec. 2000, so we have  
attributed twice the mean power measured by ACIS in 2003 to the Dec. 2000 
epoch (when Gladstone et al. \cite{Glad_02} only had HRC data, without 
spectral information with which to characterise the spectrum properly).
Emitted powers for the {\it XMM-Newton} observations of Jupiter in 2003
are reported in Branduardi-Raymont et al. (\cite{BR_07a}), but were computed 
using a factor of 2$\pi$ between flux and power, so we doubled them to 
be comparable with the rest of the measurements. The 2007 {\it XMM-Newton} 
observations, which will be the subject of a future publication 
(Branduardi-Raymont et al., in preparation), took place at the time of the 
{\it New Horizon} Jupiter fly-by, and were contemporaneous with some by 
{\it Chandra}; at this epoch, close to solar minimum, the disk X-ray power 
had dimmed considerably ($\sim$120 MW). The errors associated with the 
2003 {\it XMM-Newton} {\tt mekal} model normalisations are of the order of 
10\% (Branduardi-Raymont et al. \cite{BR_07a}), so we attributed the same 
fractional error to the corresponding emitted powers, and to the {\it Chandra}
measurements for which no errors are quoted. All these errors are contained
within the size of the plotting symbols. Larger errors ($+$100\%,$-$50\%)
apply to the 1994 {\it ROSAT} power (Waite et al. \cite{Waite_97}). The linear 
correlation coefficients are 0.60 and 0.31 for the Saturn and Jupiter datasets,
respectively. The correlation in Jupiter's case becomes stronger if we 
exclude the 1994 and 2000 measurements, for which we do not have spectral 
information (the coefficient increases to 0.54). Plotted in Fig. 5 are also 
lines derived  from a linear square fit to each dataset, which reinforce the 
view of the flux trends. The correlation between Saturn's and 
Jupiter's disk X-ray powers 
and the GOES 0.5 -- 3 \AA\ (4 -- 25 keV) flux is much weaker (coefficients of 
0.39 and 0.18 respectively). This is not surprising, since the band covers 
higher energies than those at which the $\sim$0.5 keV solar spectrum, scattered
by the planet's atmosphere, peaks.

   \begin{figure}
\hspace{-5 mm}
    \includegraphics[width=9.7cm,angle=0]{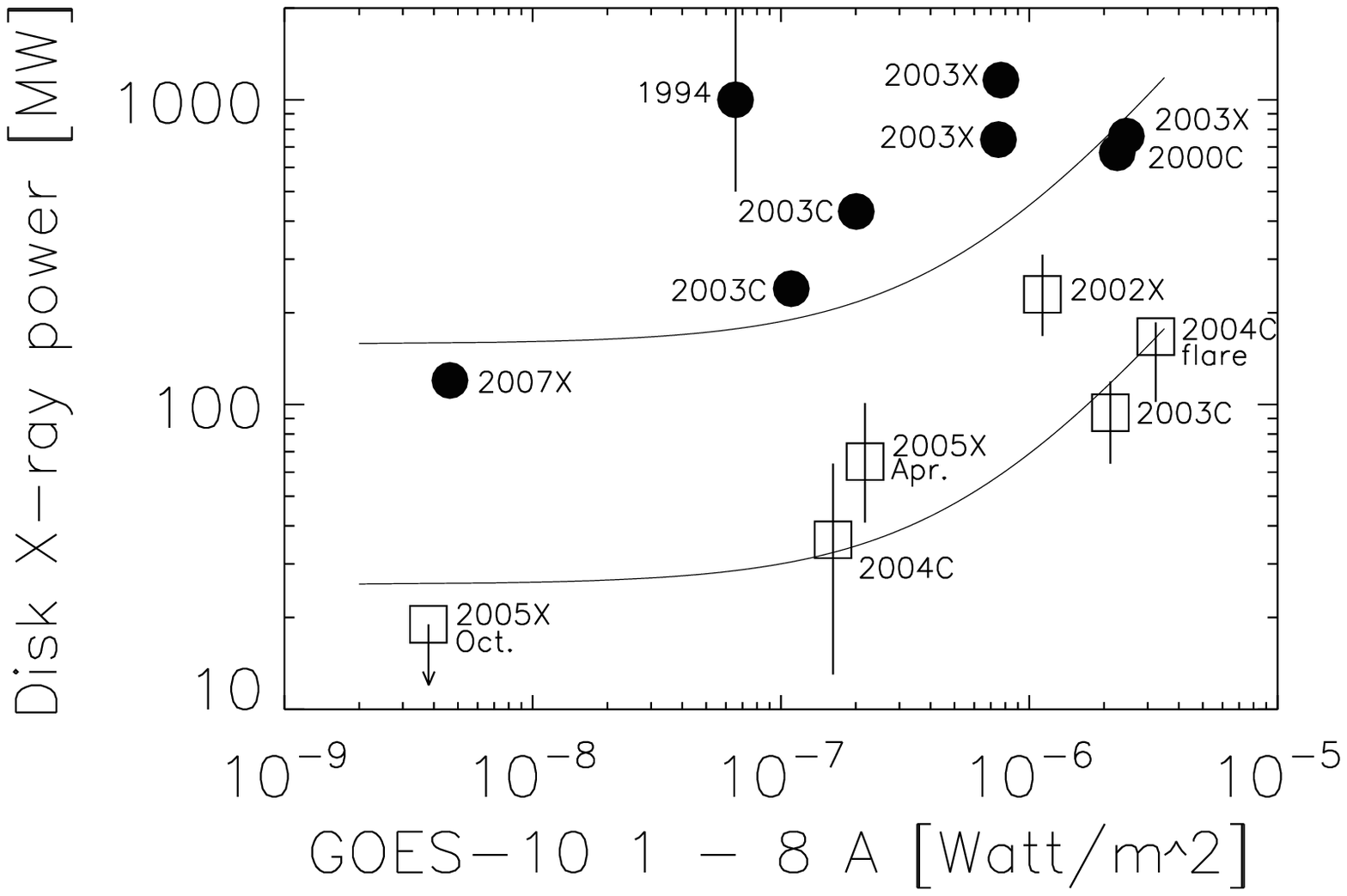}
       \caption{Saturn's (open squares) and Jupiter's (filled circles) 
                disk 0.2 $-$ 2.0 keV X-ray power plotted 
                against the solar 1 $-$ 8 $\AA$ (1.5 $-$ 12.4 keV) X-ray 
                flux measured by the GOES-10 satellite,
                for all {\it XMM-Newton} (X) and {\it Chandra} (C) 
                observations. The 1994 Jupiter data came from ROSAT. The
                continuous lines are linear square fits to each dataset.
            }
         \label{Fig5}
   \end{figure}
%

\subsection{The origin of the rings X-ray emission}

Our analysis of the complete set of Saturn's X-ray spectra over the
three year period from Oct. 2002 to Oct. 2005 suggests that the strength of 
the oxygen emission line from its rings does not vary in the same way as the 
disk emission does, following solar activity. This would be at variance 
with the interpretation of the origin of the rings emission as fluorescent 
scattering of solar X-rays from the icy rings material (Bhardwaj et al. 
\cite{Bhardwaj_05b}). However, only about one third of the ring X-ray emission 
could be explained in this way by Bhardwaj et al. (\cite{Bhardwaj_05b}).
While the fluorescence interpretation is attractive because is
underpinned by the same mechanism which leads to the disk X-ray emission,
and our spectral separation technique cannot prove that the line emission is 
truly all confined to the rings, nevertheless it is interesting to explore
a possible alternative to the solar scattering interpretation.
 
{\it Chandra} imaging in both 2003 and 2004 appears to indicate that the 
X-ray emission concentrates on the East ansa (morning side) of the rings, 
where also spokes occur; indeed Bhardwaj et al. (\cite{Bhardwaj_05b}) suggest 
that the higher X-ray brightness could be due to enhanced fluorescence on the 
ice-dust particles that are thought to make up the spokes, and to be created
by meteoritic impacts, which are more likely in the early morning hours. 
On the other hand, {\it Cassini} measurements have shown the spokes to be 
related to electrons injection (Coates et al. 2005). Recently it has also been
suggested that lightning-induced electron beams (following cosmic rays 
impacts) may produce the spokes (Jones et al. \cite{Jones_06}). Such 
electron beams may also energise the icy particles in the rings, which
could then fluoresce and cause the oxygen line emission. Thus, if the X-ray 
line is related in this way to thunderstorms that are characteristically 
'seasonal', since seasons last a long time on Saturn ($\sim$30 year sidereal 
period), one could expect the line strength to be more or less constant
over the three year period for which we have observations.

On the other hand, recent modeling work by Tseng et al. (\cite{Tseng_09})
suggests that the oxygen atmosphere of the rings is strongly dependent on 
Saturn's position in its orbit, and a much larger density is expected in
the Southern rather than in the Northern hemisphere. Thus it should not come 
as a surprise that the oxygen X-ray emission from the rings may vary over
different timescales from those of the disk emission, which is controlled by 
solar activity. Future observations may also be able to show whether the 
brightness of the oxygen X-ray emission changes after the equinox (11
August 2009).

\subsection{Are there X-ray aurorae on Saturn? and on Uranus and Neptune?}

Unlike for Jupiter (Branduardi-Raymont et al. \cite{BR_07b}), there is no
evidence of auroral X-ray emission in the Saturn data obtained so far.
The X-ray spectrum of the Jovian aurorae is made up of two components: 
the softer one is generally assumed to be produced by 
energetic heavy ions precipitating
in the planet's atmosphere and radiating line emission following charge 
exchange with atmospheric particles; the other, becoming dominant 
above 2 keV, is due to energetic electrons also precipitating through the
atmosphere and radiating bremsstrahlung. Recently, a morphological study
of the spatial distribution of this high energy emission in Jupiter's
Northern polar region has shown it to be co-located with the main FUV
auroral oval and bright FUV regions (Branduardi-Raymont et al. \cite{BR_08}).
This, and the relative emission powers in the FUV and hard X-rays, support the 
view that on Jupiter these two emissions are produced by the same population 
of energetic electrons precipitating through the atmosphere. Given the bright 
and variable FUV aurorae observed on Saturn, one would expect some correlation 
with the X-ray emission to be present.

On the basis of model calculations by Singhal et al. (\cite{Sing_92}) and 
Waite et al. (\cite{Waite_92}), Branduardi-Raymont et al. (\cite{BR_08}) 
deduce that the power in the FUV emission is expected to be some 10$^{5}$ to 
10$^{6}$ times that in hard X-rays if the same electrons are producing both
and conclude that the observed emissions from Jupiter's aurorae (i.e. 
45 MW in hard X-rays and 340 GW in FUV) are broadly 
consistent with these ratios. We can make similar considerations for Saturn,
which also possesses a dense atmosphere, like Jupiter's:
restricting the analysis to the more sensitive {\it XMM-Newton} observations,
we extract 3$\sigma$ upper limits in the band 2.0 $-$ 8.0 keV of 16, 35 and 37
counts for datasets (1), (2) and (3), respectively. Adopting a bremsstrahlung
temperature of 10 keV (lower than the 90 keV best fitting the Jupiter spectra
of Branduardi-Raymont et al. \cite{BR_07b}, since the precipitating electrons 
are generally of lower energy on Saturn than on Jupiter, Gustin et al. 
\cite{Gustin_09}, G\'{e}rard et al \cite{Gerard_09}), and 
using the EPIC-pn response matrices to determine the conversion 
factor from countrate to energy flux, we find corresponding 3$\sigma$ upper
limits to the emitted 2.0 $-$ 8.0 keV power of 147, 94 and 107 MW. The 
{\it XMM-Newton} (3) observation (28 Oct. 2005, 06:34 to 29 Oct. 2005, 
11:32 UT) was contemporaneus with three exposures of the Saturnian aurora
taken with the HST Advanced Camera for Surveys, during
a coordinated campaign with in-situ measurements by the {\it Cassini} 
spacecraft. This was a time of quiet solar wind conditions, and yet the UV 
aurora was seen to be very variable; the measured UV power at the time of the 
{\it XMM-Newton} observation varied between 3.3 and 8.0 GW (G\'{e}rard et al. 
\cite{Gerard_06}). If we apply to Saturn 
the ratios of 10$^{5}$ to 10$^{6}$ between hard X-ray and FUV powers calculated
for Jupiter, it is clear that any electron bremsstrahlung emission would have 
been some 3 to 4 orders of magnitude fainter than our current {\it XMM-Newton}
upper limits.
 
To make a further comparison of the most magnetised planets in the solar 
system, in Table 3 we summarise parameters that can be 
expected to be relevant to the auroral characteristics for all the outer 
planets, as well as Earth. With respect to Jupiter, Saturn is down by a 
factor of $\sim$ 20 and 30 in surface magnetic field and magnetic moment  
respectively. A similar ratio applies to the maximum average FUV aurora
input power. Moreover, other parameters are likely to matter in the aurora
production, such as the magnetospheric particle density: this is significantly 
higher, by at least another factor of 30, around Jupiter (being fed by Io's 
volcanoes) than Saturn (where it is sustained by the rings and the icy moons).
Thus, if Saturn's total auroral X-ray power (including both soft X-rays of 
ionic origin and the higher energy electron bremsstrahlung) is lower than that 
of Jupiter by the combination of the magnetic field and particle density 
factors (around three orders of magnitude), it may measure only a few MW; this 
is equivalent to an X-ray flux at Earth of a few $\times$ 10$^{\rm -17}$ erg 
cm$^{\rm -2}$ s$^{-1}$, so it is no surprise that it is going undetected 
with current instrumentation. In fact, Saturn's aurora may be at the limit 
of detectability even with the next
generation most powerful International X-ray Observatory (IXO), being planned 
by ESA, NASA and JAXA, which is expected to have a sensitivity limit of 
$\sim$ 10$^{\rm -17}$ erg cm$^{\rm -2}$ s$^{-1}$.

One may argue that our scaling approach is rather simplistic, but there are
other examples in the planetary context where it seems to be applicable.
Radio events reported by Louarn et al. (\cite{Lou_07}) at both Jupiter and 
Saturn suggest that plasma transport in the magnetosphere and the associated
release of rotational energy may operate in a similar way at the two planets, 
but scale according to different timescales for internal processes. This
is also in line with the conclusions of Stallard et al. (2008) that the 
differences in auroral emissions between the two giant outer planets are 
determined by scaling differences in their physical conditions and 
environments, while the processes leading to aurora formation are essentially
the same (e.g. breakdown of co-rotation). There are terrestrial analogues to
the scaling idea as well: solar wind disturbances at Earth are larger in 
amplitude and faster (order of minutes) than at Jupiter and Saturn (timescales 
of hours) because of the smaller magnetosphere and the lack of steady supply 
of plasma (Clarke et al. \cite{Clarke_09} and references therein).

Assuming that the same considerations are valid for Uranus and Neptune, and 
scaling on their dipole magnetic moments, which are down by a factor of 400 
and 800 from Jupiter respectively, we would expect FUV auroral strengths of 
the order of 100 to 50 GW in the two cases. However, the magnetospheric 
particle density at these planets is some 10$^{\rm 3}$ times lower than at 
Jupiter (Table 3), suggesting that we should expect much lower FUV emitted 
powers than this initial extrapolation implies. The observed FUV powers, 
reported by Bhardwaj and Gladstone (\cite{BG_00}), are at most 40 and 0.1 GW: 
the latter value is of the order which we would expect for the low plasma 
content 
of Neptune's magnetosphere, but Uranus does not seem to scale down in the 
same way; perhaps the very large tilt (59$^{\rm o}$) between its magnetic 
dipole and rotation axis and the 30\% offset of the dipole from the planet's 
centre, causing an order of magnitude difference in surface magnetic field 
strength between day- and night-side, have also the effect of enhancing the 
power output in the aurora. In any case, taking these estimates together 
with the vaste distances from Earth of these two planets excludes the 
possibility of detecting their X-ray auroral emissions with current 
instruments. Indeed Bhardwaj et al. (\cite{Bhardwaj_07}) report an 
unsuccessful attempt by {\it Chandra} to detect X-rays from Uranus
in August 2002. These authors also suggest that Uranus and Neptune,
as well as Saturn's moon Titan, are likely to shine in X-rays, both by 
particle precipitation in the magnetosphere, and by scattering of solar 
X-rays. Looking to the future, let us consider the most optimistic case 
for Uranus, at about 4 times the distance from Earth than Jupiter, with an FUV 
output $\sim$ 1000 times lower than Jupiter: ignoring the much smaller 
magnetospheric particle density, and scaling from the observed Jovian 
0.2 - 2 keV auroral emission (few times 10$^{\rm -14}$ erg cm$^{\rm -2}$ 
s$^{-1}$, Branduardi-Raymont et al. \cite{BR_07b}) implies a flux at Earth 
from Uranus of $\sim$ 10$^{\rm -18}$ erg cm$^{\rm -2}$ s$^{-1}$, far  
too low for current X-ray observatories, and about an order of magnitude below
IXO's expected sensitivity limit. 

\section{Conclusions}

   \begin{enumerate}
      \item Statement 1. We report the results of all the X-ray observations of
Saturn carried out to date. We present a detailed spectral analysis of the 
available data which reinforces the earlier conclusion (Bhardwaj et al. 
\cite{Bhardwaj_05a}) that the planet's disk acts as a reflector of solar 
X-rays, much in the same way as observed for Jupiter, Earth, Venus and Mars 
(cf. Bhardwaj et al. \cite{Bhardwaj_07}). 
      \item Statement 2. Our results 
also suggest that the strength of the X-ray emission from the Saturnian rings, 
in the form of a fluorescence oxygen line, does not systematically decrease, 
like that from the disk, while the solar activity cycle evolves towards 
its minimum. As a possible
alternative to scattering of solar X-rays in the rings icy material, we 
suggest that the origin of the line may be related to the formation of the 
spokes, which appear on the East ansa of the rings as do the bright X-ray 
spots seen by {\it Chandra}, by lightning-induced electron beams.
      \item Statement 3. Unlike Jupiter and Earth, we do not find evidence for 
X-ray aurorae on Saturn. This can be explained by the limited sensitivity of 
the observations to date, if the auroral X-ray and FUV emissions on Saturn 
have relative strengths similar to those on Jupiter. Other factors, such as 
magnetic field strength, magnetospheric particle densities and internal plasma
sources may also contribute.    
      \item Statement 4. After comparison with Jupiter and Earth, we conclude 
that Saturn is unique in its X-ray characteristics as it is in the
FUV (Clarke et al. 2005). We still have much to learn about the precise 
mechanisms operating in its aurorae, and in its rings, leading (or not, 
as it may be) to X-ray 
production. The effectiveness of further observations with {\it Chandra} and 
{\it XMM-Newton} will increase as we move out of the current (2009) minimum in 
the solar activity cycle, providing invaluable additional information on the 
evolution of both, disk and rings emissions, and of their response to the solar
X-ray flux and the solar wind. However, a much more efficient way to pursue
and enhance these studies, and relate the electromagnetic output of the 
planet to particle and magnetic field measurements, would involve flying 
an X-ray spectrometer to operate in-situ, onboard a spacecraft mission to 
the Saturnian system. We look forward to the possible realisation of this 
project in association with the missions to the outer planets currently being
studied by both ESA and NASA. Such a mission to Saturn is most likely to 
involve a close encounter with Titan, and an X-ray spectrometer would
return unique data on this enigmatic satellite, and its interactions with
Saturn and the solar wind as well, especially at times when it is outside 
Saturn's magnetosphere. In-situ X-ray observations are also the 
only feasible way to search for auroral X-ray emission from Uranus and 
Neptune.

   \end{enumerate}

\begin{acknowledgements}
      GBR acknowledges useful discussions with G. H. Jones and suggestions
      by T. E. Cravens.
      This work is based in part on observations obtained with 
      {\it XMM-Newton},
      an ESA science mission with instruments and contributions directly
      funded by ESA Member States and the USA (NASA). The Mullard Space 
      Science Laboratory acknowledges financial support from the UK 
      Science and Technology Facilities Council.  
\end{acknowledgements}

\end{document}